\documentclass[onecolumn,prd,nofootinbib,preprintnumbers]{revtex4}
\usepackage{epsfig}
\usepackage{graphicx,xcolor,amsmath}

\newcommand{\be}{\begin{equation}}
\newcommand{\ee}{\end{equation}}
\newcommand{\bea}{\begin{eqnarray}}
\newcommand{\eea}{\end{eqnarray}}

\begin{document}
\preprint{LAPTH-018/21}
\title{First implications of Tibet AS$_\gamma$ data for heavy dark matter}
\author{Arman Esmaili}
\email{arman@puc-rio.br}
\affiliation{Departamento de F\'isica, Pontif\'icia Universidade Cat\'olica do Rio de Janeiro, Rio de Janeiro 22452-970, Brazil}
\author{P. D. Serpico}
\email{serpico@lapth.cnrs.fr}
\affiliation{Univ. Grenoble Alpes, Univ. Savoie Mont Blanc, CNRS, LAPTh, F-74940 Annecy, France}

\begin{abstract} 
Extensive air shower detectors of gamma rays in the sub-PeV energy region provide a new and relatively unexplored window for dark matter searches. Here we derive some implications of the recently published Tibet AS$_\gamma$ data for decaying dark matter candidates. The available spectral information is already useful in obtaining competitive constraints, surpassing existing limits above 10 PeV mass for hadronic or massive boson final states. This is particularly true if accounting for a benchmark astrophysical background of Galactic cosmic rays in the (0.1-1) PeV range. By relying on the arrival distribution of the photons, we show that significantly better sensitivity can be attained, comparable or better than IceCube also for most leptonic final states. Full data exploitation requires however further information disclosure. 
\end{abstract}
\maketitle
\date{\today}

\section{Introduction}

The last decade has seen the opening of two windows in astrophysics: The high-energy neutrino one~\cite{Aartsen:2013bka,Aartsen:2013jdh} and the gravitational wave one~\cite{Abbott:2016blz}. As it is usually the case,  these seminal discoveries have defied some of the expectations and have stimulated new fascinating questions. For instance, the origin of the bulk of IceCube events is still a puzzle, with mounting indications that they could involve new classes of objects, either astrophysical or exotic ones (for recent studies in that sense, see~\cite{Capanema:2020oet,Capanema:2020rjj}). Decaying dark matter (DM) has long been recognized as one of these possibilities~\cite{Feldstein:2013kka,Esmaili:2013gha}, capable at the same time to account for the peculiar energy and angular distributions of the events~\cite{Esmaili:2014rma}\footnote{For further studies of heavy DM signatures in neutrino/gamma-ray experiments see \cite{Esmaili:2012us,Murase:2015gea,Esmaili:2015xpa,Dev:2016qbd,Chianese:2016kpu,Kalashev:2017ijd,Bhattacharya:2017jaw,Chianese:2017nwe,Abeysekara:2017jxs,Neronov:2018ibl,Sui:2018bbh,Kachelriess:2018rty,Chianese:2019kyl,Ishiwata:2019aet,Dekker:2019gpe,Neronov:2020wir}.}. In~\cite{Esmaili:2015xpa}, we made the point that extensive air shower (EAS) detectors sensitive to sub-PeV gamma-rays could provide amazing sensitivity to this class of DM models, either supporting such an interpretation of IceCube data or significantly improving the constraints. We forecasted that the forthcoming generation of observatories would attain the needed sensitivity for crucial studies. With the first detection of Galactic gamma-rays up to PeV energies by Tibet AS$_\gamma$~\cite{Amenomori:2021gmk} and the commissioning of the LHAASO observatory~\cite{He:2020ipy}, we are witnessing the dawn of this new astrophysical window, and should prepare to harness its full potential, in particular in a multiwavelength and multimessenger context (see~\cite{Fang:2021ylv,Liu:2021lxk,Qiao:2021iua} for initial examples of this approach). In this article, we provide an assessment of the importance of this first detection for DM searches, a task whose relevance has already been recognized in~\cite{Dzhatdoev:2021xjh}. This energy range has its own peculiarities, notably the  absence of any extragalactic contribution (since the gamma-rays are fully absorbed on the CMB and IR background, cascading down to sub-TeV energies) and the need to take into account partial and anisotropic absorption of the gamma-rays even within the Galaxy, as detailed in~\cite{Esmaili:2015xpa}. In Sec.~\ref{spec}, we briefly introduce the dataset and the model used, deriving spectral constraints. Sec.~\ref{ang} illustrates the power of the angular analysis, whose full exploitation will have to wait for the disclosure of further experimental information. In Sec.~\ref{concl}, we report our conclusions. For the first time in this context, we also include a model of Galactic diffuse gamma-rays according to~\cite{Lipari:2018gzn} to assess the impact of astrophysical backgrounds on DM searches in this window. Due to the very rudimentary understanding of the CR sources and propagation properties in this regime, this has to be seen as a motivation for a better assessment of these predictions and of their uncertainties. 

\section{\label{spec}Spectral analysis}

The Tibet air shower and muon detector array has detected for the first time a sub-PeV gamma-ray flux from the Galaxy~\cite{Amenomori:2021gmk}. The collaboration reports energy spectra in two angular windows in Galactic coordinates: Region I ( $|b|<5^\circ$ and $25^\circ<l<100^\circ$) overlaps with the region from which the ARGO-YBJ collaboration reported a diffuse detection in the TeV range~\cite{Bartoli:2015era}; Region II ($|b|<5^\circ$ and $50^\circ<l<200^\circ$) matches the one where CASA-MIA previously only reported upper limits~\cite{Borione:1997fy}. DM bounds derived from both regions are comparable, with Region II marginally better (at the 10\% level) in some regions of parameter space. Bounds from ARGO-YBJ data in Region I kick in and dominate at sub-PeV masses. Below, for each DM decay channel, the best limit from Regions I and II and of either experiment is reported at each value of DM mass.

For the DM signal template, we consider the DM decay models discussed in~\cite{Bhattacharya:2019ucd}, which we address the reader to for further details. For the spectra of gamma-rays and $e^\pm$ from DM decay we use the results of~\cite{Bauer:2020jay}. Anisotropic absorption of the gamma-rays in Galaxy via $e^\pm$ production on IR/optical/UV photons (plus the isotropic absorption onto CMB) is accounted for as in~\cite{Esmaili:2015xpa}. Also, the inverse-Compton scattering of $e^\pm$ from DM decay off the ambient photons (mostly CMB photons) has been taken into account (for technical aspects,  see~\cite{Esmaili:2015xpa}), although this contribution remains $\lesssim$ a few percent not only in the Galactic plane region, but also for $|b|<40^\circ$, which is considered in Sec.~\ref{ang}. For the DM halo we use the Navarro-Frenk-White density profile~\cite{Navarro:1996gj} with the critical radius $\simeq24$~kpc and Sun-GC distance of $8.3$~kpc, yielding the DM density at Solar System $\simeq0.39~{\rm GeV}~{\rm cm}^{-3}$ (for further details see section 3 of~\cite{Esmaili:2015xpa}).

 \begin{figure}[t]
{\includegraphics[width=0.32\textwidth]{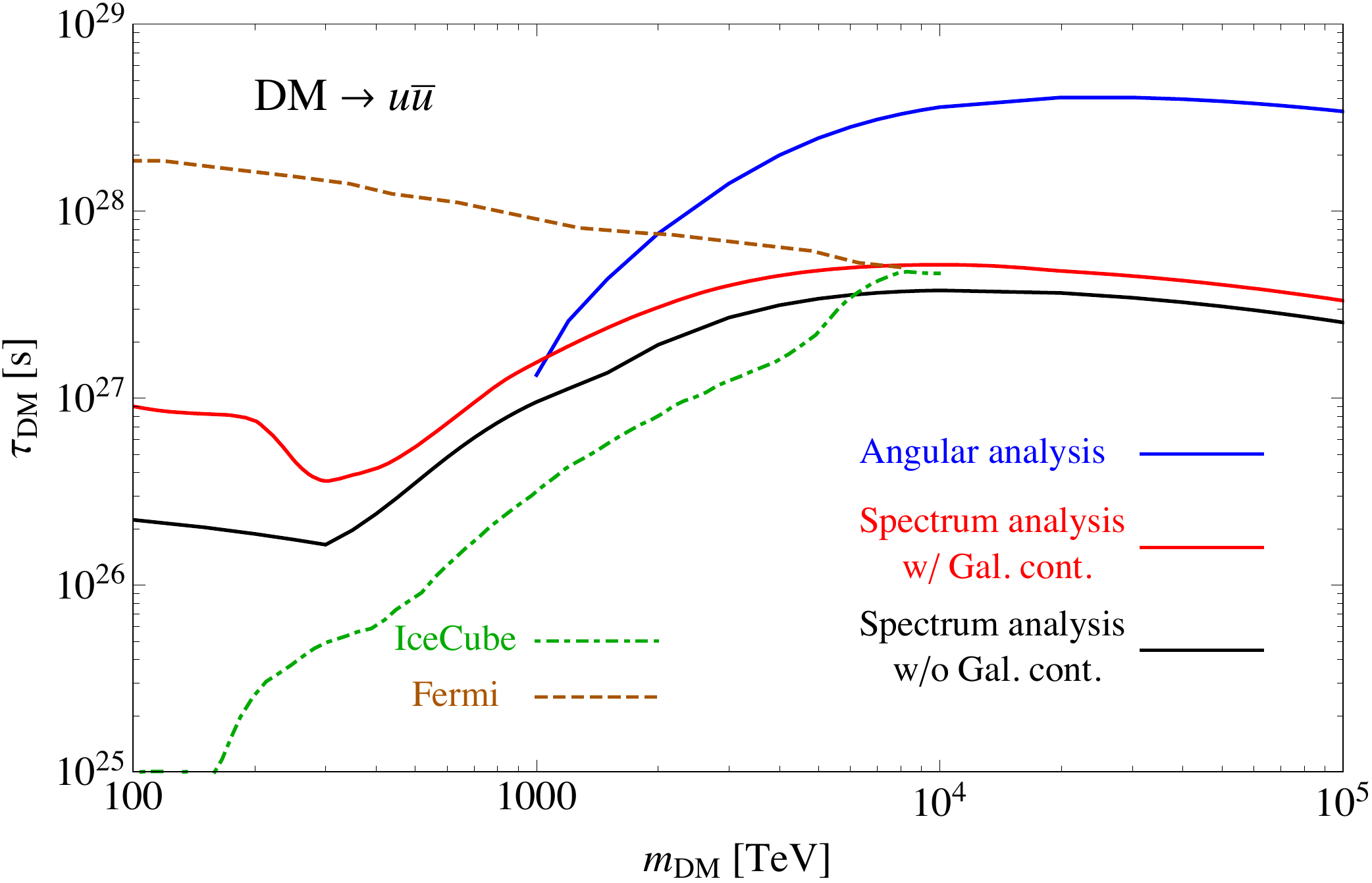}}
{\includegraphics[width=0.32\textwidth]{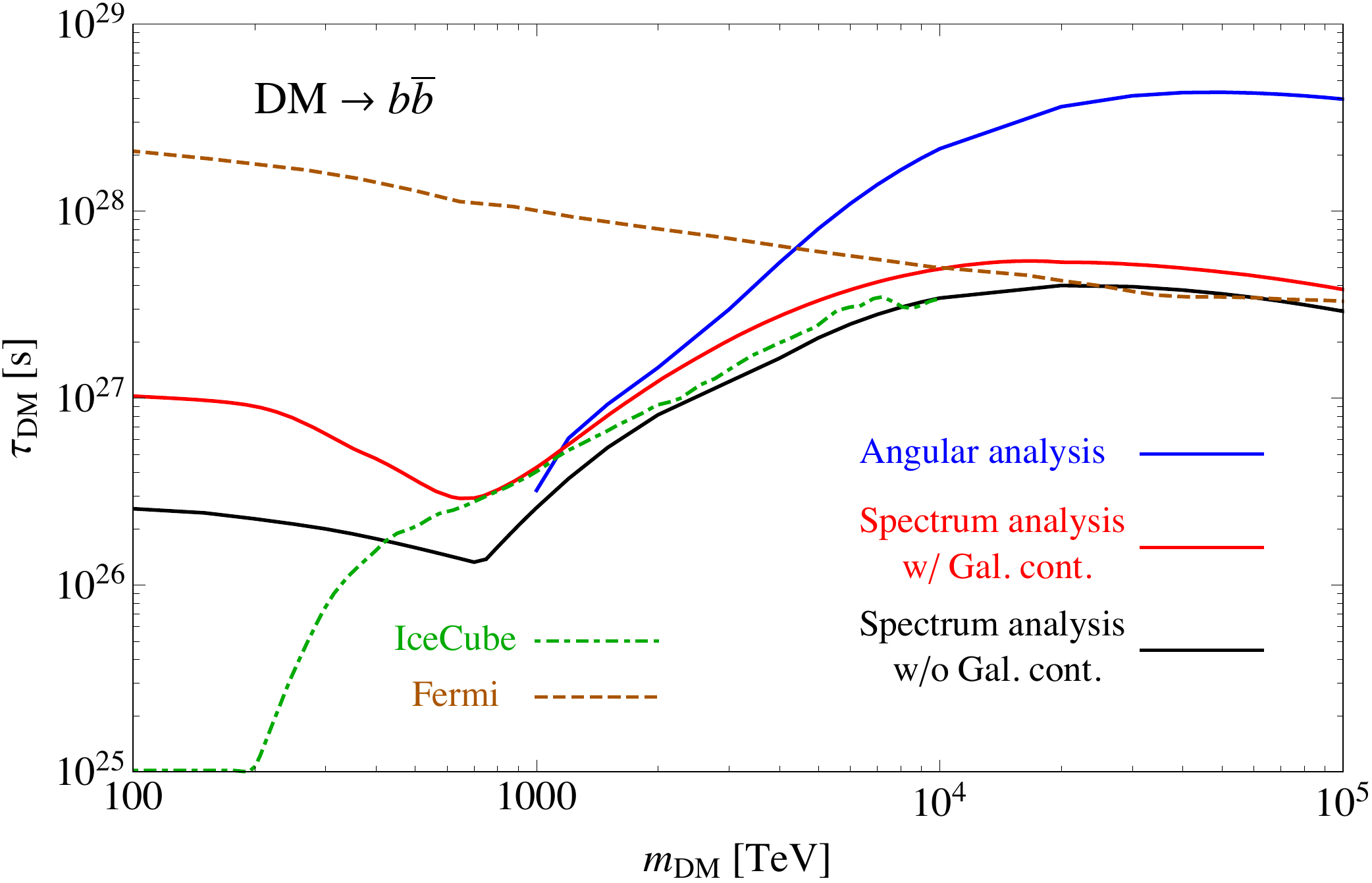}}
{\includegraphics[width=0.32\textwidth]{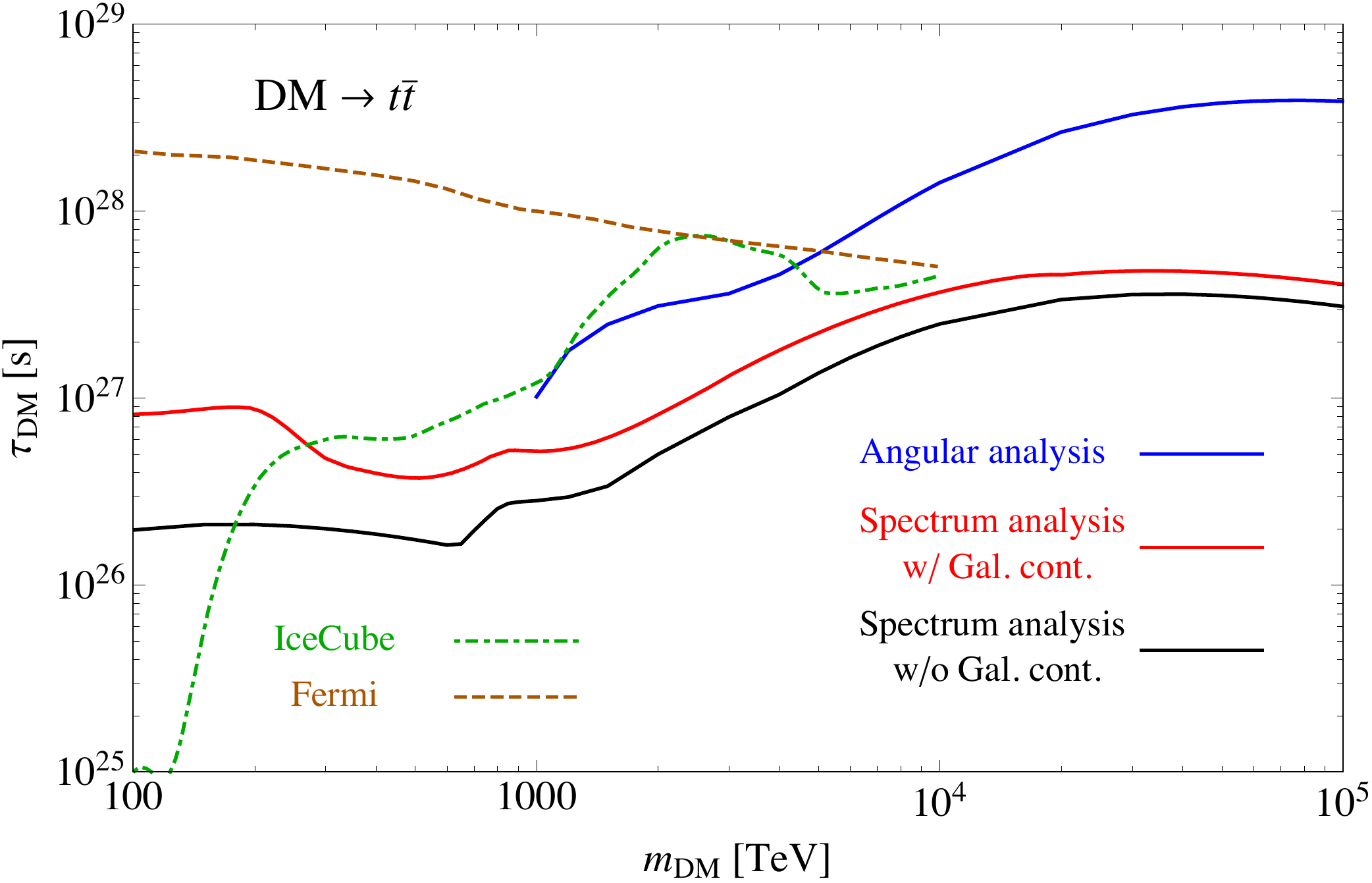}}\\
{\includegraphics[width=0.32\textwidth]{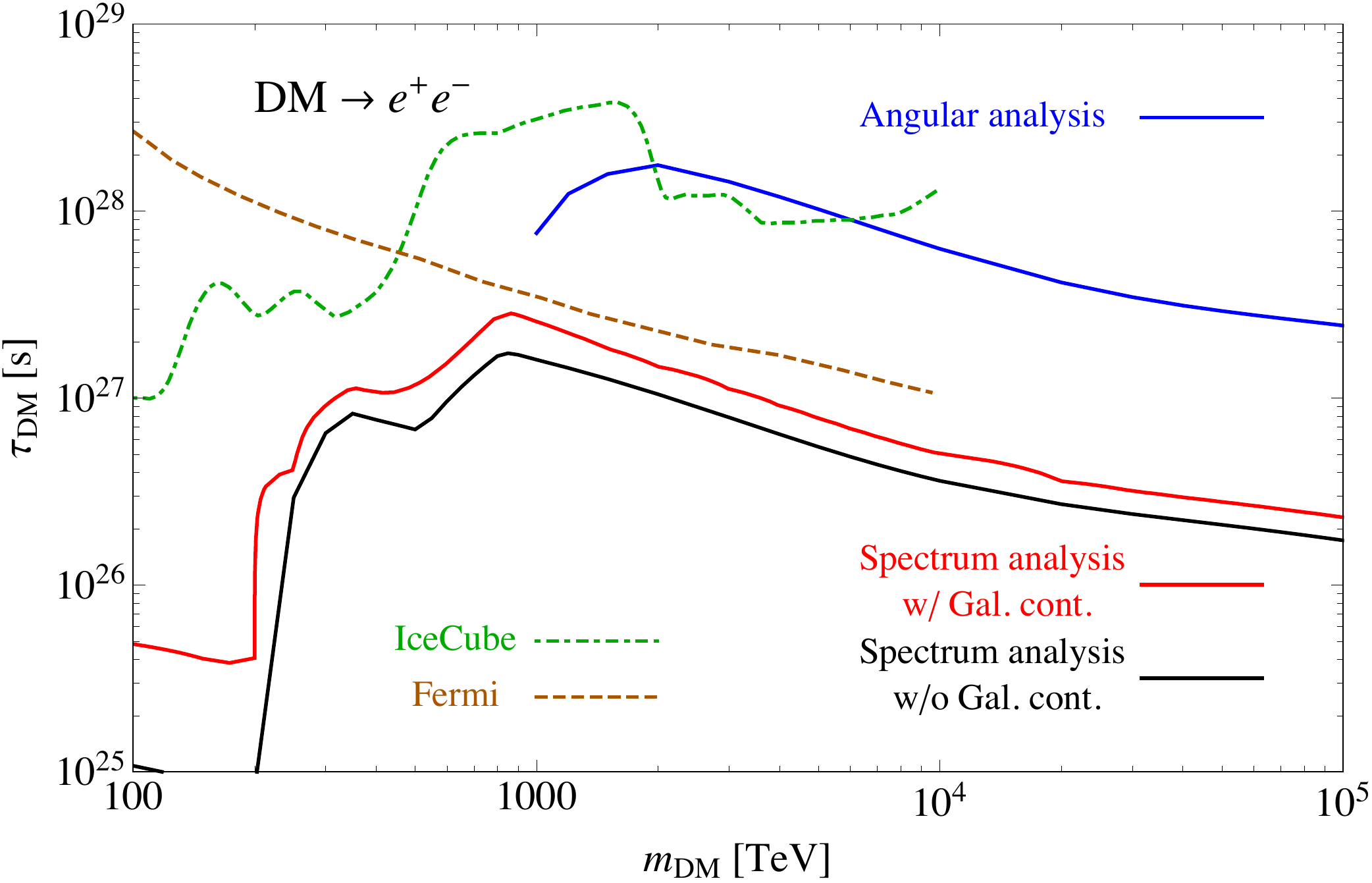}}
{\includegraphics[width=0.32\textwidth]{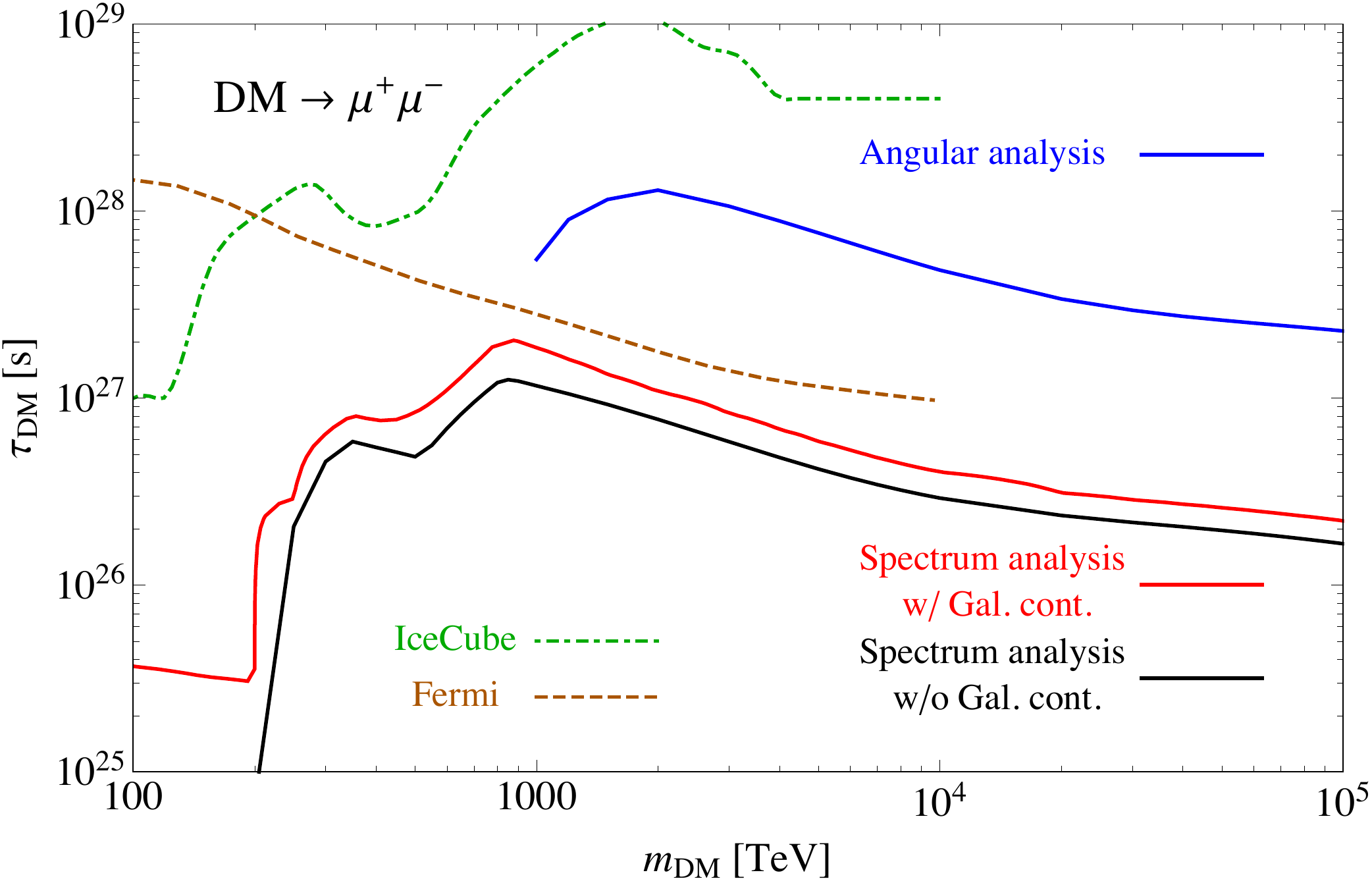}}
{\includegraphics[width=0.32\textwidth]{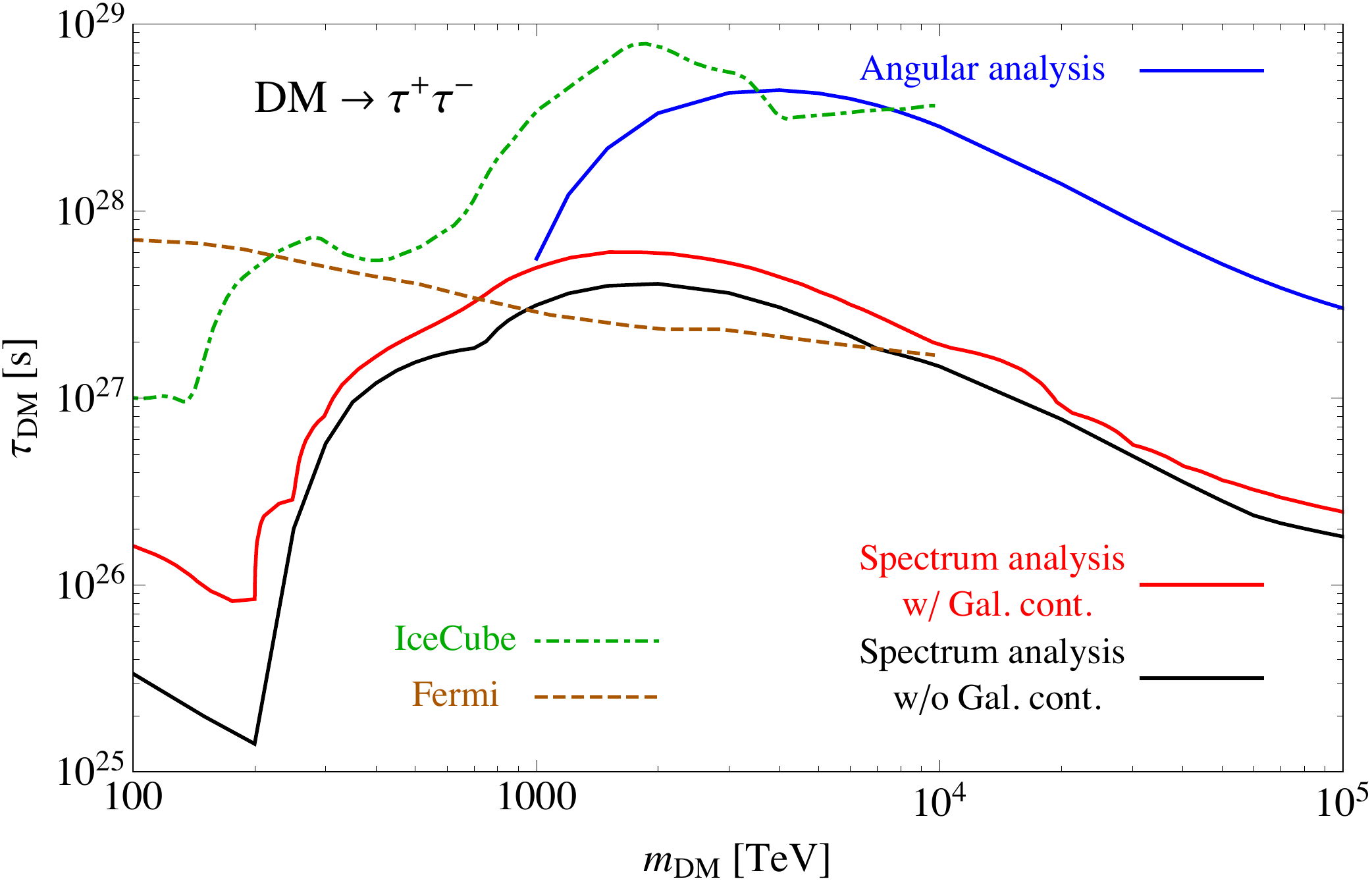}}\\
{\includegraphics[width=0.32\textwidth]{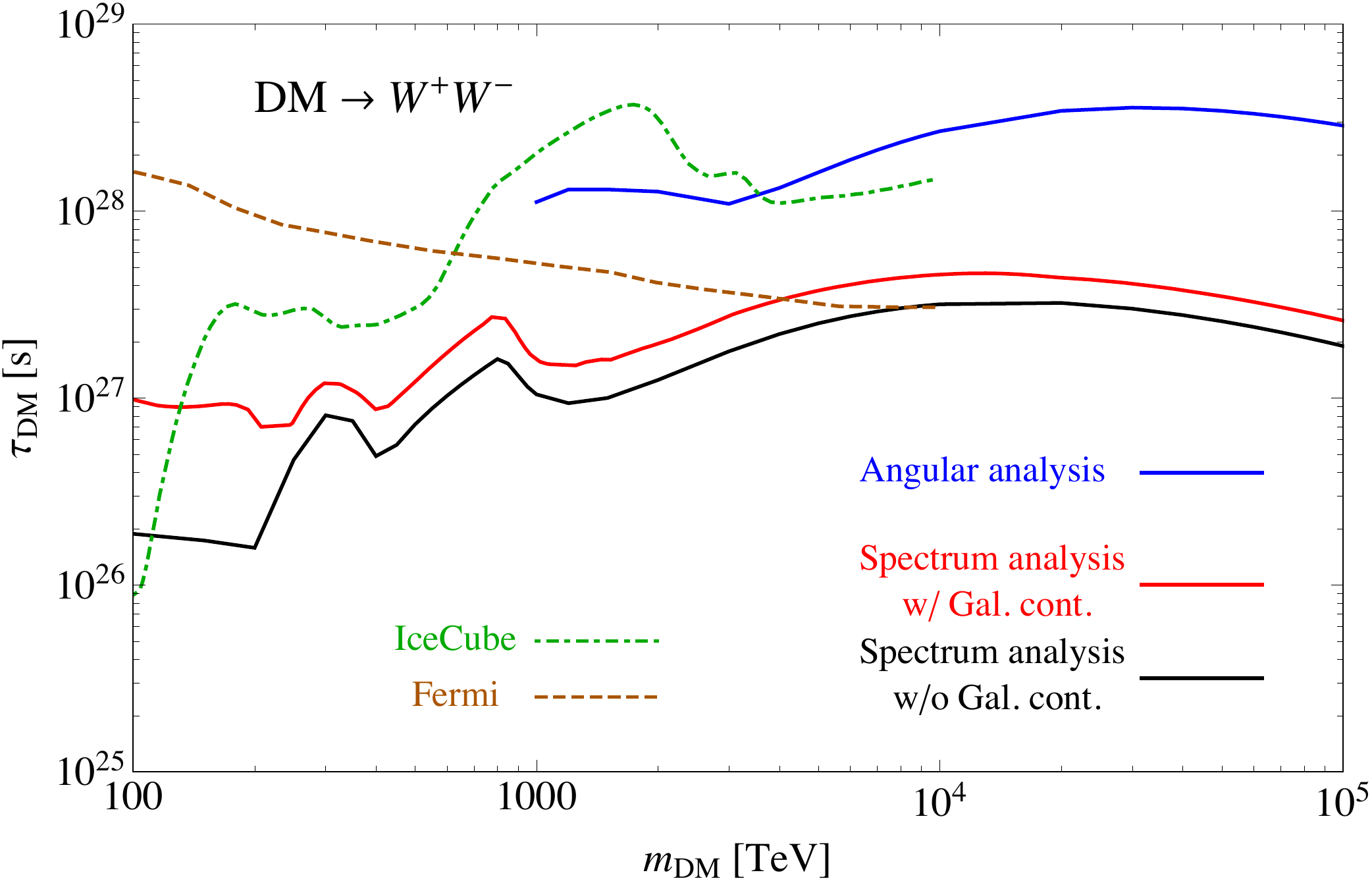}}
{\includegraphics[width=0.32\textwidth]{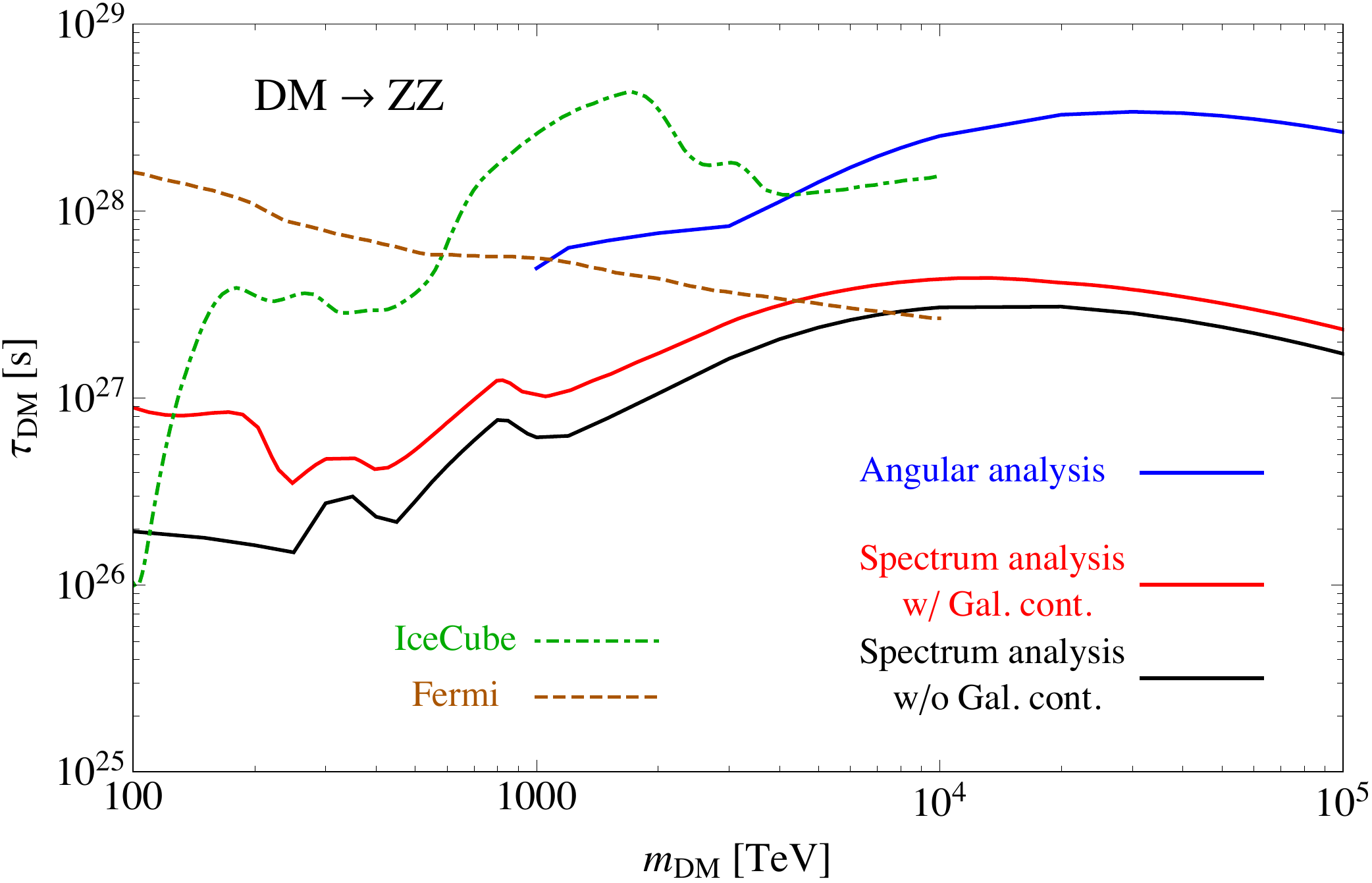}}
{\includegraphics[width=0.32\textwidth]{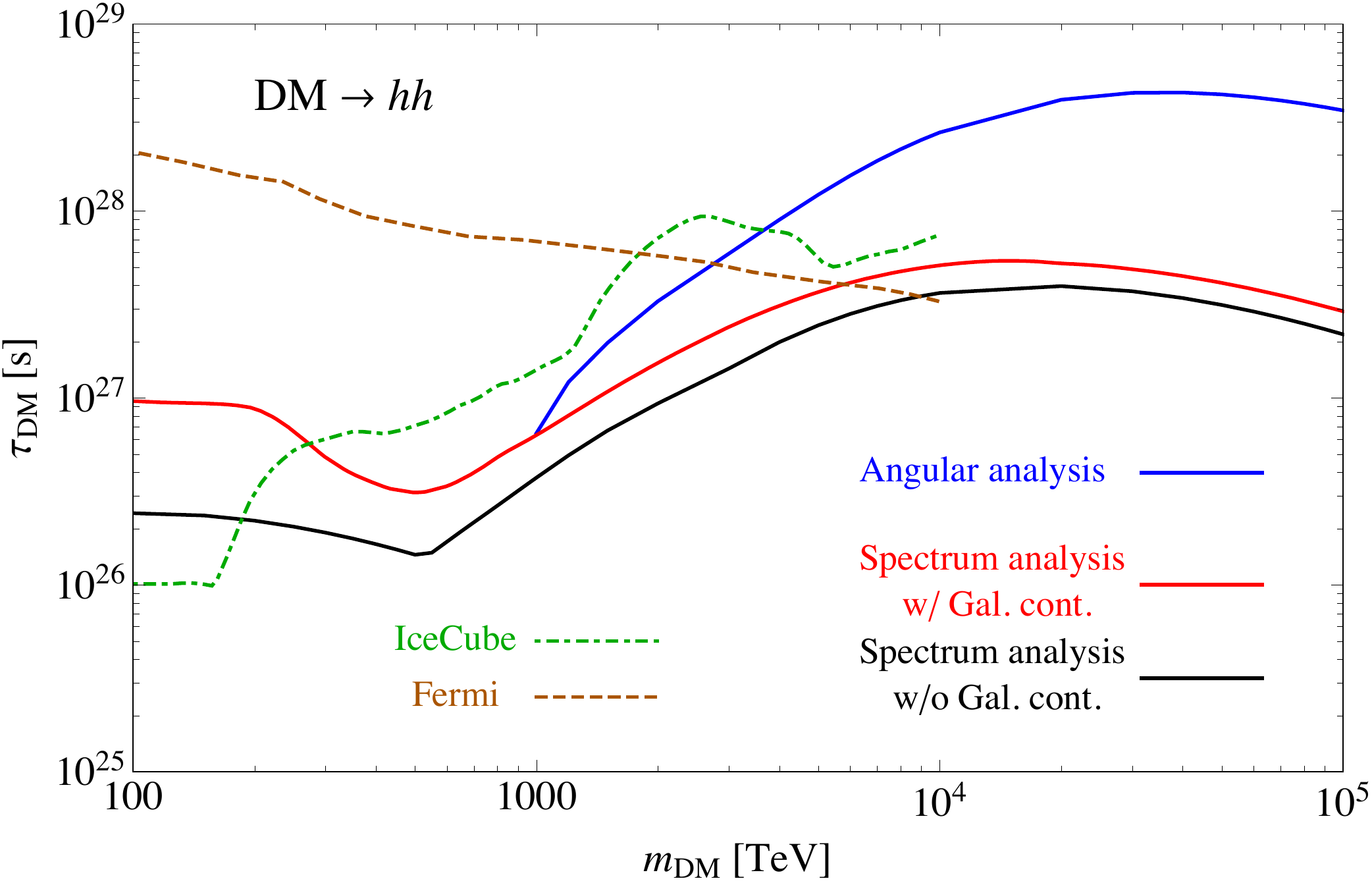}}\\
{\includegraphics[width=0.32\textwidth]{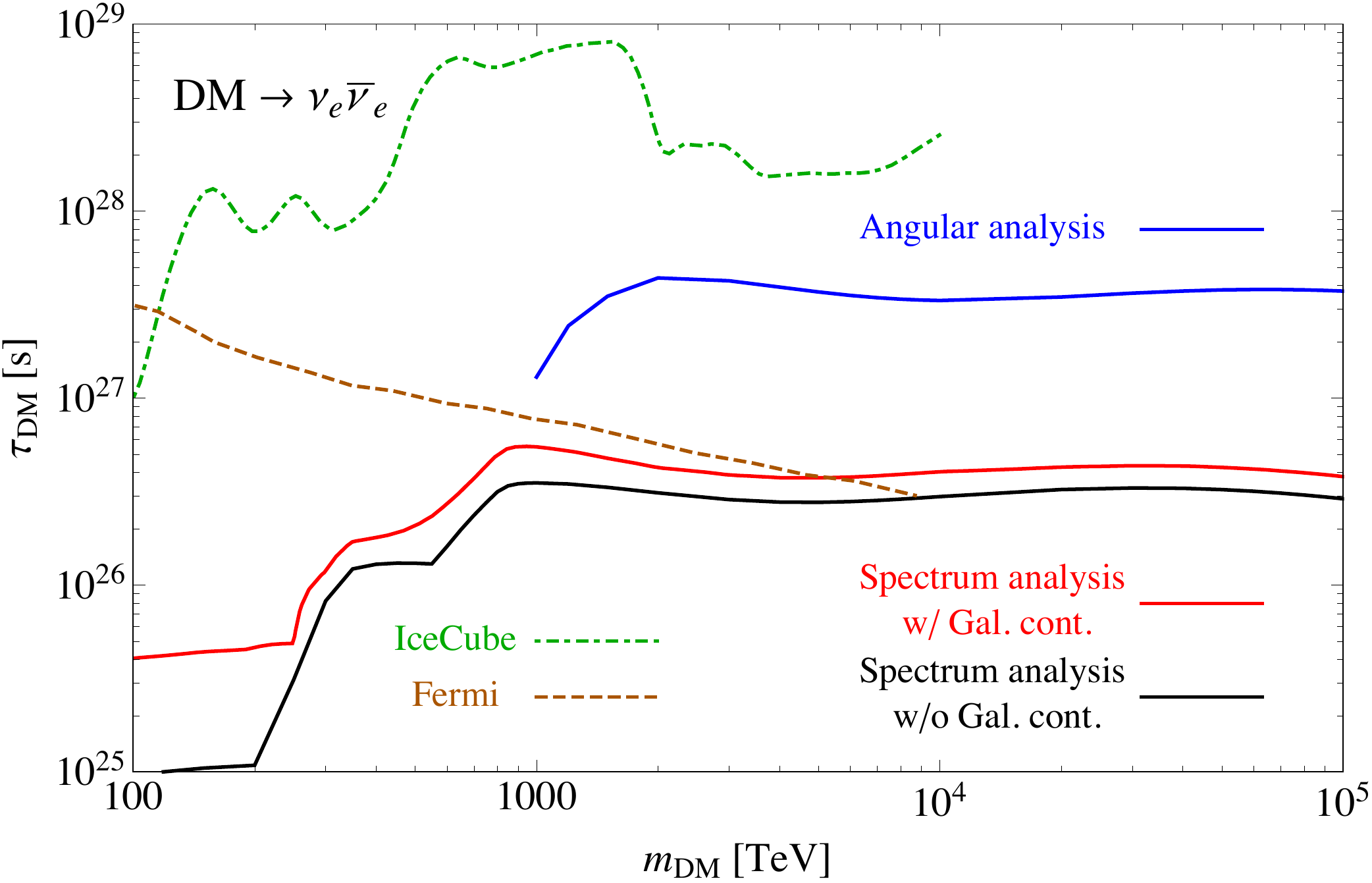}}
{\includegraphics[width=0.32\textwidth]{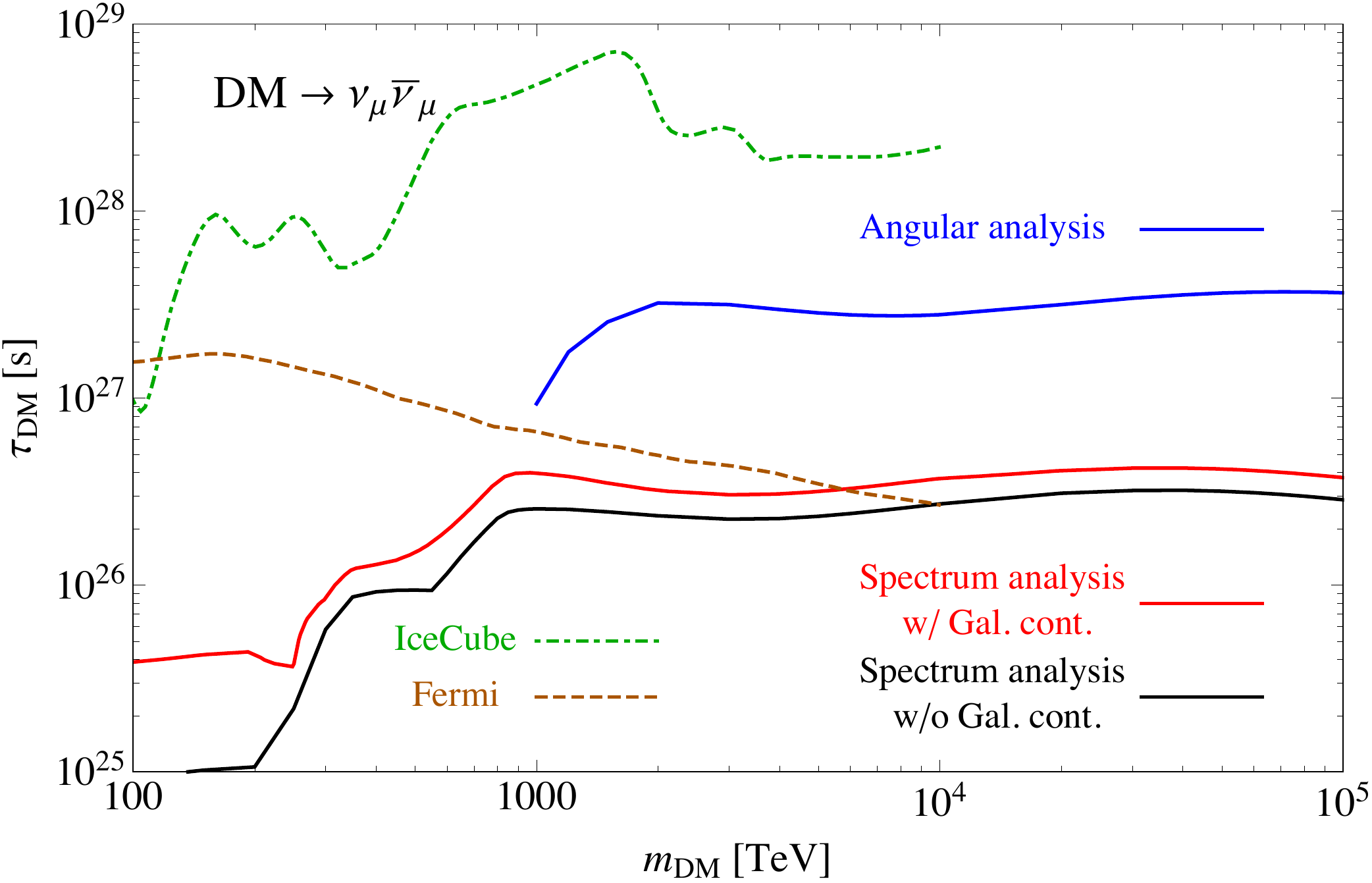}}
{\includegraphics[width=0.32\textwidth]{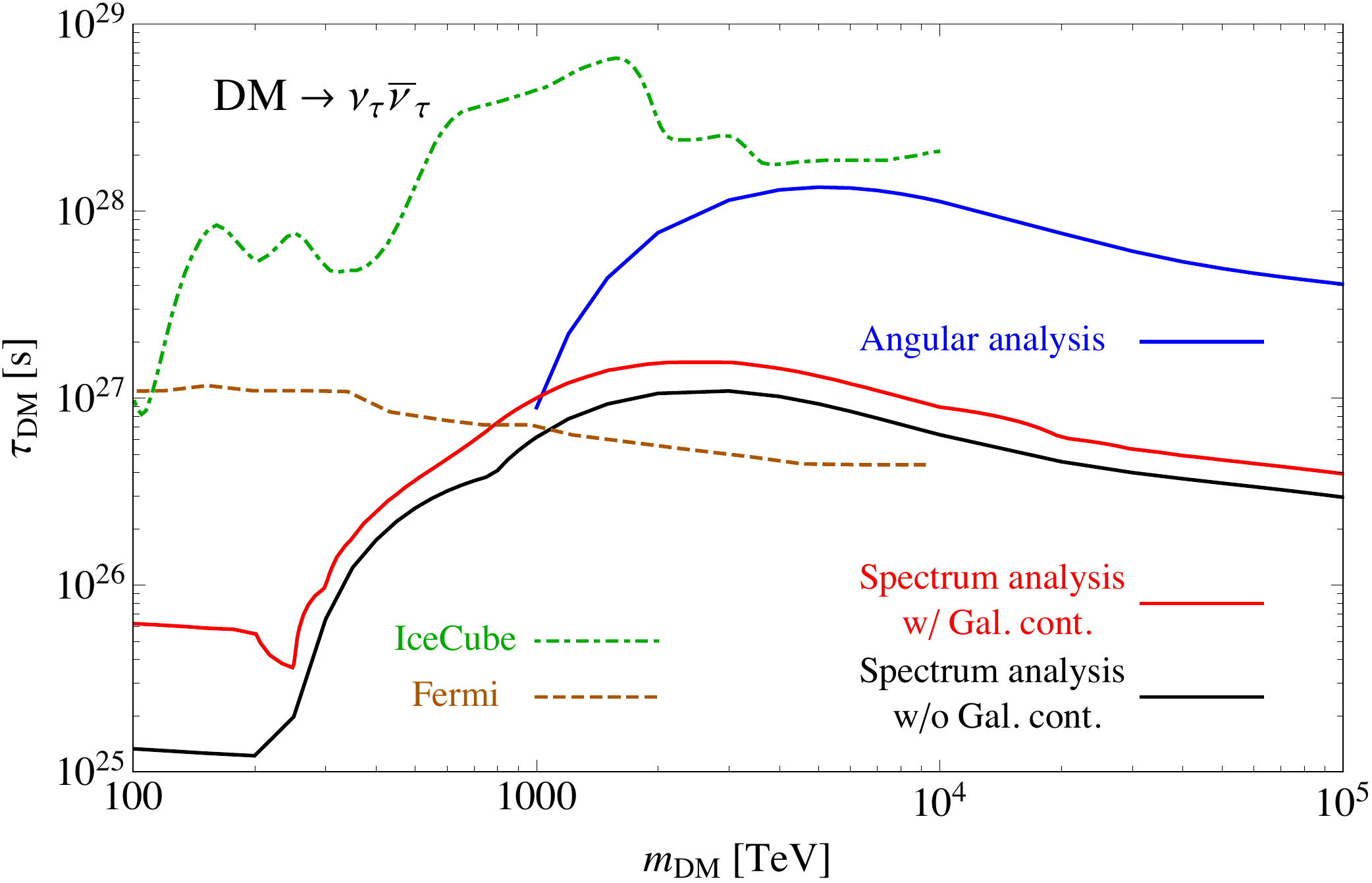}}
	\caption{Bounds on the DM lifetime for decay channels DM $\to u\bar{u},\, b\bar{b}, \, t\bar{t}$ (top row, left to right); DM $\to e^+e^-,\, \mu^+\mu^-,\, \tau^+\tau^-$ (second row, left to right); DM $\to W^+W^-, \,ZZ,\, hh$ (third row, left to right);  DM $\to \nu_e\bar{\nu}_e,\, \nu_\mu\bar{\nu}_\mu,\, \nu_\tau\bar{\nu}_\tau$  (bottom row, left to right). The blue and black solid curves show the tentative bounds from angular and spectral analyses (without background), respectively (see text for methodological explanations); the red curves show the 95\% C.L. bounds from the spectral analysis including the Galactic contribution from~\cite{Lipari:2018gzn}. The brown dashed and green dot-dashed curves show the 95\% C.L. bounds from IceCube~\cite{Bhattacharya:2019ucd} and Fermi-LAT~\cite{Cohen:2016uyg}, respectively.}
	\label{fig:allch}
\end{figure}

The bounds derived from the spectral analysis of Tibet AS$_\gamma$ data,  reported in three energy bins $100<E_\gamma/{\rm TeV}<158$,  $158<E_\gamma/{\rm TeV}<398$ and $398<E_\gamma/{\rm TeV}<1000$, are presented in Fig.~\ref{fig:allch} for twelve  decay channels. The black solid curves in Figure~\ref{fig:allch} show the bounds derived simply by requiring the $\gamma$-ray flux from DM to remain below the $1\sigma$ upper limits reported by Tibet AS$_\gamma$ in all the three energy bins. The upturns in sub-PeV DM masses are the consequence of ARGO-YBJ data at $\sim1$~TeV. Inclusion of the ``space-independent'' astrophysical model of diffuse gamma-rays from ref.~\cite{Lipari:2018gzn}, leads to stronger exclusion. The red solid curves in Fig.~\ref{fig:allch} show the 95\% C.L. bounds obtained from $\chi^2$ analysis that includes this Galactic contribution. For DM masses $m_{\rm DM}\lesssim1$~PeV, the bounds from spectral analysis are weaker than the limits obtained from diffuse gamma-ray data of Fermi-LAT~\cite{Cohen:2016uyg} for almost all the decay channels. For some channels, like the $u$ and $b$-quark, the spectral analysis bounds improve over the IceCube's limits~\cite{Bhattacharya:2019ucd} at $m_{\rm DM}\lesssim1$~PeV. Note that these bounds are based on spectra extracted from Pythia's results~(for details, see~\cite{Sjostrand:2019zhc}), while ours rely on~\cite{Bauer:2020jay}: We checked that differences for masses below $\mathcal{O}(10)$~PeV are typically at the few percent level, hence irrelevant for our purposes.

The newly derived spectral limits surpass Fermi-LAT background limits for masses above $\sim 10$~PeV for all hadronic final state channels, as well as massive gauge-bosons and Higgs. While for light leptons they do not improve over Fermi, they do for third generation leptons already at the PeV scale, although they are not yet competitive with IceCube. It is worth noting that, for light or $b$-quark final states, Tibet AS$_\gamma$'s spectral bounds surpass both Fermi and IceCube bounds above $\sim 10\,$PeV, in particular if the astrophysical background is accounted for. 

\section{\label{ang}Angular distribution analysis}

\begin{figure}[!th]
{\includegraphics[width=0.49\textwidth]{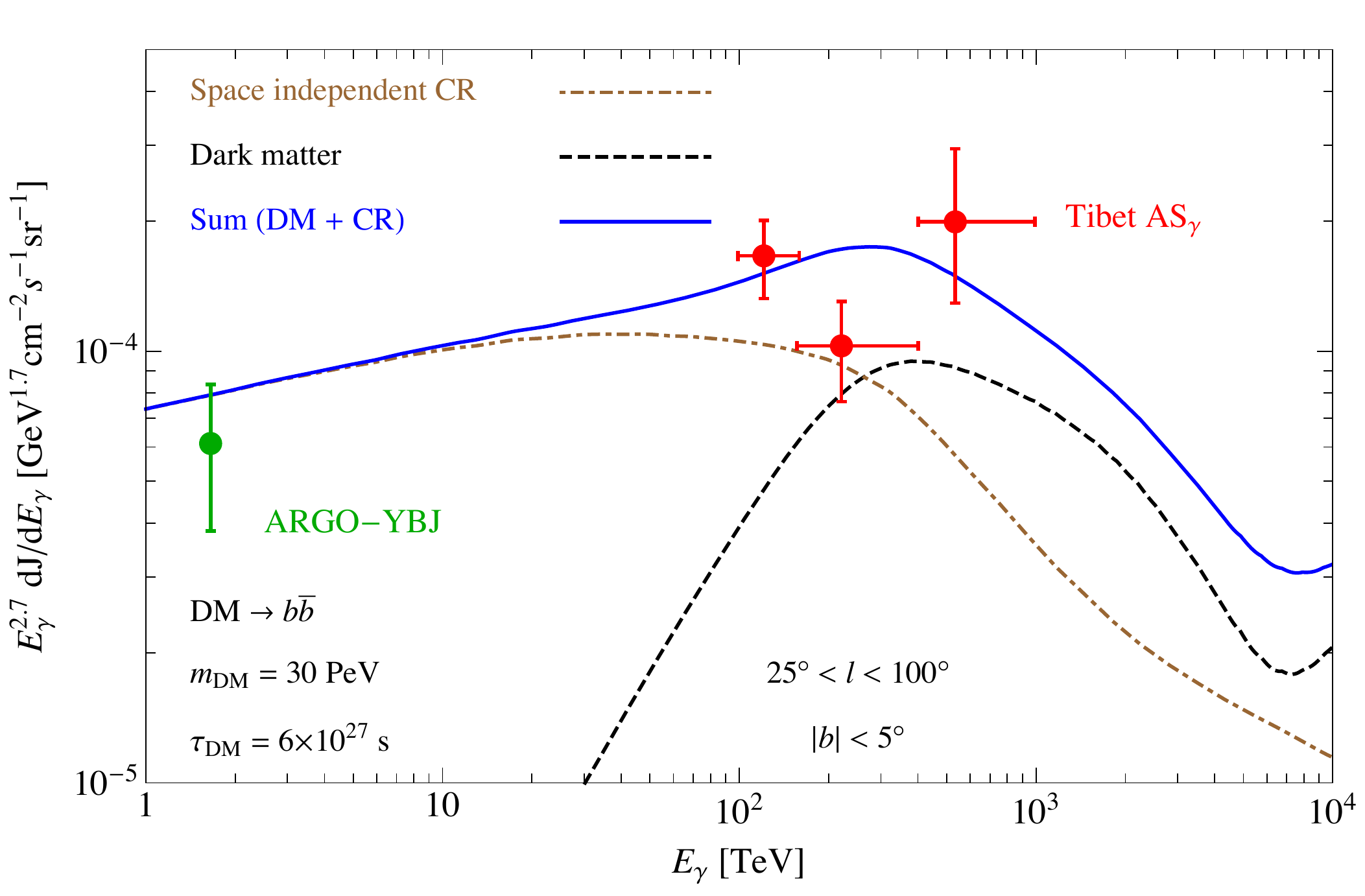}}
{\includegraphics[width=0.48\textwidth]{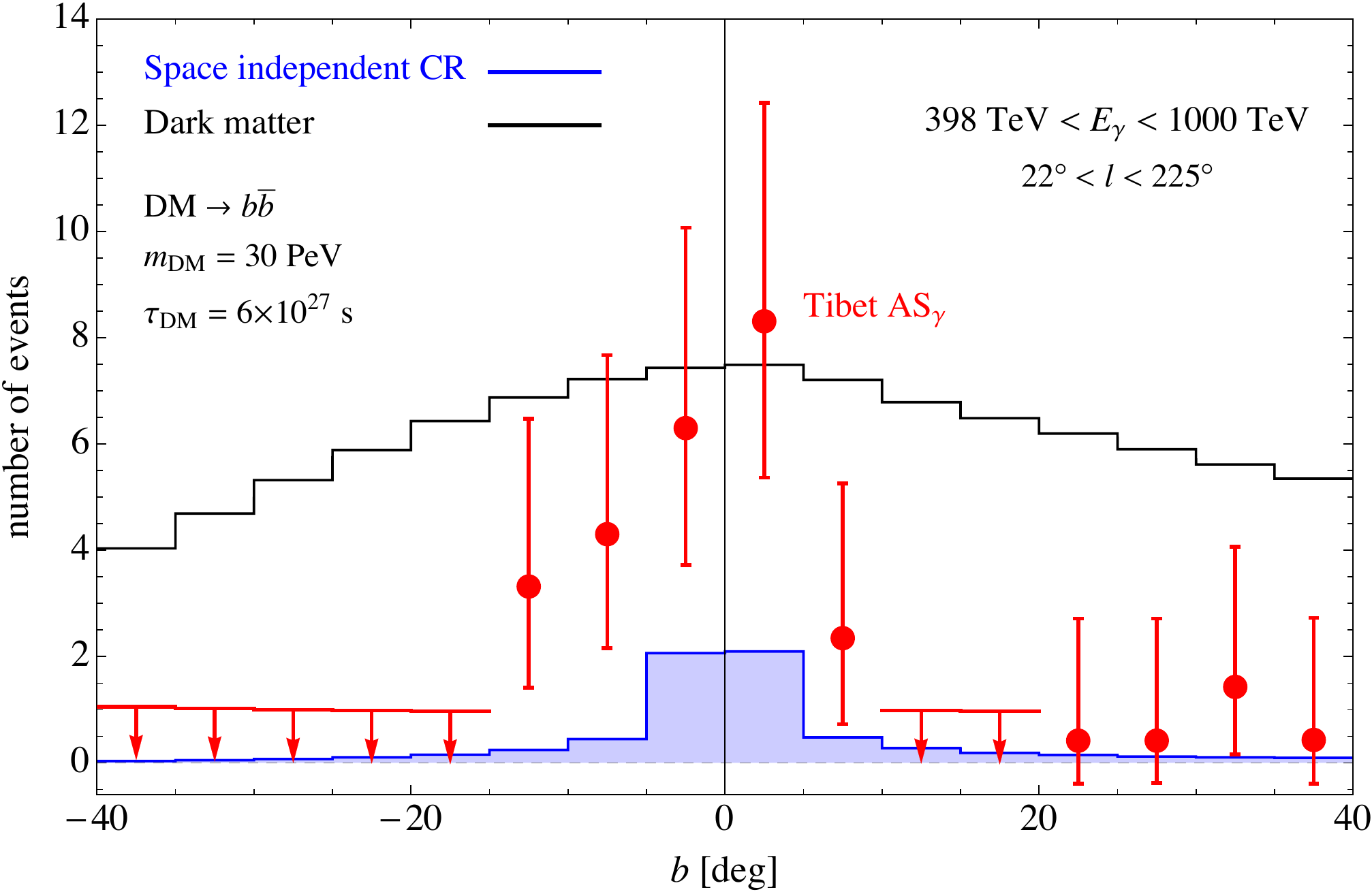}}
\caption{Left panel: the Tibet AS$_\gamma$ (red points) and ARGO-YBJ (green point) data, and the theoretical spectra associated to the Galactic contribution (brown dot-dashed curve), DM (black dashed curve) and the sum (blue solid curve), for Region I. For the DM we assume $m_{\rm DM}=30$~PeV, decaying into $b\bar{b}$ with the lifetime $\tau_{\rm DM}=6\times10^{27}$~s. Right panel: the latitude profile of the observed events (red points, with $1\sigma$ error bars), the expected Galactic contribution (blue histogram) and the DM model of left panel.}\label{model}
\end{figure}

Interestingly, the measured diffuse gamma-ray flux by Tibet AS$_\gamma$ and reported in~\cite{Amenomori:2021gmk} is compatible with an excess with respect to the existing evaluation of Galactic diffuse gamma-ray flux of~\cite{Lipari:2018gzn}, in particular in Region II which is farther away from the inner Galaxy. This is qualitatively in agreement with expectations from decaying DM models, whose flux declines very gently at directions away from the Galactic Center. Merely based on spectral information, an excess could admit a DM decay interpretation, as illustrated in the left panel of Fig.~\ref{model}: The black dashed curve shows the gamma-ray flux from DM $\to b\bar{b}$ with the mass and lifetime $(m_{\rm DM},\tau_{\rm DM})=(30~{\rm PeV},6\times10^{27}~{\rm s})$ which is at the edge of exclusion (see the red solid curve in the middle panel at the top row of Fig.~\ref{fig:allch}), and is compatible with the limits from IceCube and Fermi-LAT. The blue solid curve, showing the sum of DM and Galactic contributions, provides a better fit to the Tibet AS$_\gamma$ data than solely the Galactic contribution. The DM model in the left panel of Fig.~\ref{model} is only one example among various spectrally viable DM models, particularly if one allows DM decay into multiple channels.    

Angular information of Tibet AS$_\gamma$'s data provides an exquisite handle to probe a wider range of DM models and test the viability of those that may even appear favoured by spectral considerations only and is exemplified in the left panel of Fig.~\ref{model}. Unfortunately, a  complete angular analysis is not possible at the moment: Although the collaboration reports in~\cite{Amenomori:2021gmk} the Galactic latitude profiles of data in the three energy bins, the use of these data is made difficult by the lack of precise information on the actual experimental aperture (or exact exposure) and $\gamma$-efficiency, especially its energy dependence. Also, the authors of~\cite{Amenomori:2021gmk} renormalize (separately for each energy bin) the Galactic model of~\cite{Lipari:2018gzn} to the observed number of events within $|b|<5^\circ$, which prevents inferring the detector's characteristics. In addition, the background estimation at high  Galactic latitudes needs clarification and likely improvement. The  method currently employed in~\cite{Amenomori:2021gmk} for the first two energy bins ($100<E_\gamma/{\rm TeV}<158$ and $158<E_\gamma/{\rm TeV}<398$)  yields $1\sigma$ upper limits on the number of events (or excess of events, subtracting the estimated background from observed events) which are negative already at $|b|\sim10^\circ$. Taking these upper limits at face value, no model (either conventional Galactic models or DM decay ones) would survive. As a consequence of these limitations, here we just illustrate how powerful such an analysis could be, by using the Galactic latitude distribution of observed events in the third energy bin $398<E_\gamma/{\rm TeV}<1000$. The right panel of Fig.~\ref{model} shows the latitude profile of events in the highest energy bin. The black curve is the expectation of the DM model considered in the left panel and the red points show the Tibet AS$_\gamma$ data with $1\sigma$ error bars. In the calculation of the latitude profile for DM we assume that the acceptance efficiency of the detector depends on declination (and is uniform with respect to right ascension) according to the geometrical arguments of~\cite{Sommers:2000us}. The blue shaded histogram shows the expected number of events from the Galactic model of~\cite{Lipari:2018gzn} rescaled back to its true normalization by a factor equal to the ratio of the averaged Galactic flux to the measured flux
. Clearly, the angular distribution of events at $|b|>5^\circ$ in the highest energy bin, without considering the Galactic model, excludes the DM model of the left panel in Fig.~\ref{model}.

Following the crude application of not-overshooting the $1\sigma$ upper limits on angular profile of observed events in the highest energy bin (without taking into account the Galactic contribution), the blue curves in Fig.~\ref{fig:allch} represent our tentative bounds from angular considerations. {\it For hadronic final states, as well as massive gauge and Higgs bosons, Tibet AS$_\gamma$ data constitute the most sensitive probe of decaying DM above a few PeV. Even for leptonic final states, they become comparable or better than IceCube at high DM masses, and noticeably better than Fermi-LAT limits.} Once the angular distribution of events in the lower energy bins could be properly analyzed, we anticipate a significant improvement in sensitivity also at lower DM masses. Proper availability of these data may be also useful to probe other large-scale astrophysical Galactic features~\cite{Ahlers:2013xia}.

\section{\label{concl}Conclusions}

In this article we have presented a first examination of the impact of the recent detection of Galactic gamma-rays up to PeV energies by Tibet AS$_\gamma$~\cite{Amenomori:2021gmk} for heavy decaying DM models. Our main conclusions are that the spectral data alone improve over previous EAS bounds,  making them competitive (i.e. sometimes better than best existing ones) above 10 PeV or so, particularly for hadronic or massive boson final states. This is particularly true if the astrophysical Galactic background is included. The angular distribution of the data holds an even greater potential. We illustrated how the angular analysis would make Tibet AS$_\gamma$ competitive or better than IceCube even for leptonic final states which generally produce more neutrinos, and undoubtedly better than Fermi-LAT bounds. 

Our two main recommendations are: 
i) To the Tibet AS$_\gamma$ collaboration, to disclose more information in order for the sensitivity power of the angular analysis to be fully exploited. In particular, detailed information of the detector exposure and its latitude dependence (in case of difference with respect to the geometrical corrections) and the energy-dependence of $\gamma$-efficiency are essential. Also, an improved estimation of background events at high latitudes is crucial for the viability of angular analysis in lower energy bins. ii) To the cosmic ray theory community, to refine our understanding of the astrophysical background at PeV energies and of its uncertainties, which may prove instrumental also to tighten bounds by typically 30\% or so, according to the current benchmark.

\begin{acknowledgments}
This research used the computing resources and assistance of the John David Rogers Computing Center (CCJDR) in the Institute of Physics ``Gleb Wataghin'', University of Campinas.
\end{acknowledgments}


\begin{thebibliography}{00}

\bibitem{Aartsen:2013bka}
M.~G.~Aartsen \textit{et al.} [IceCube],
Phys. Rev. Lett. \textbf{111}, 021103 (2013)
[arXiv:1304.5356 [astro-ph.HE]].

\bibitem{Aartsen:2013jdh}
M.~G.~Aartsen \textit{et al.} [IceCube],
Science \textbf{342}, 1242856 (2013)
[arXiv:1311.5238 [astro-ph.HE]].


\bibitem{Abbott:2016blz}
B.~P.~Abbott \textit{et al.} [LIGO Scientific and Virgo],
Phys. Rev. Lett. \textbf{116}, no.6, 061102 (2016)
[arXiv:1602.03837 [gr-qc]].

\bibitem{Capanema:2020oet}
A.~Capanema, A.~Esmaili and P.~D.~Serpico,
JCAP \textbf{02}, 037 (2021)
[arXiv:2007.07911 [hep-ph]].

\bibitem{Capanema:2020rjj}
A.~Capanema, A.~Esmaili and K.~Murase,
Phys. Rev. D \textbf{101}, no.10, 103012 (2020)
[arXiv:2002.07192 [hep-ph]].

\bibitem{Feldstein:2013kka}
B.~Feldstein, A.~Kusenko, S.~Matsumoto and T.~T.~Yanagida,
Phys. Rev. D \textbf{88}, no.1, 015004 (2013)
[arXiv:1303.7320 [hep-ph]].

\bibitem{Esmaili:2013gha}
A.~Esmaili and P.~D.~Serpico,
JCAP \textbf{11}, 054 (2013)
[arXiv:1308.1105 [hep-ph]].

\bibitem{Esmaili:2014rma}
A.~Esmaili, S.~K.~Kang and P.~D.~Serpico,
JCAP \textbf{12}, 054 (2014)
[arXiv:1410.5979 [hep-ph]].

\bibitem{Esmaili:2012us}
A.~Esmaili, A.~Ibarra and O.~L.~G.~Peres,
JCAP \textbf{11}, 034 (2012)
[arXiv:1205.5281 [hep-ph]].

\bibitem{Murase:2015gea}
K.~Murase, R.~Laha, S.~Ando and M.~Ahlers,
Phys. Rev. Lett. \textbf{115}, no.7, 071301 (2015)
[arXiv:1503.04663 [hep-ph]].

\bibitem{Esmaili:2015xpa}
A.~Esmaili and P.~D.~Serpico,
JCAP \textbf{10}, 014 (2015)
[arXiv:1505.06486 [hep-ph]].

\bibitem{Dev:2016qbd}
P.~S.~B.~Dev, D.~Kazanas, R.~N.~Mohapatra, V.~L.~Teplitz and Y.~Zhang,
JCAP \textbf{08}, 034 (2016)
[arXiv:1606.04517 [hep-ph]].


\bibitem{Chianese:2016kpu}
M.~Chianese, G.~Miele and S.~Morisi,
JCAP \textbf{01}, 007 (2017)
[arXiv:1610.04612 [hep-ph]].

\bibitem{Kalashev:2017ijd}
O.~E.~Kalashev and M.~Y.~Kuznetsov,
JETP Lett. \textbf{106}, no.2, 73-80 (2017)
[arXiv:1704.05300 [astro-ph.HE]].

\bibitem{Bhattacharya:2017jaw}
A.~Bhattacharya, A.~Esmaili, S.~Palomares-Ruiz and I.~Sarcevic,
JCAP \textbf{07}, 027 (2017)
[arXiv:1706.05746 [hep-ph]].

\bibitem{Chianese:2017nwe}
M.~Chianese, G.~Miele and S.~Morisi,
Phys. Lett. B \textbf{773}, 591-595 (2017)
[arXiv:1707.05241 [hep-ph]].

\bibitem{Abeysekara:2017jxs}
A.~U.~Abeysekara \textit{et al.} [HAWC],
JCAP \textbf{02}, 049 (2018)
[arXiv:1710.10288 [astro-ph.HE]].

\bibitem{Neronov:2018ibl}
A.~Neronov, M.~Kachelrie\ss{} and D.~V.~Semikoz,
Phys. Rev. D \textbf{98}, no.2, 023004 (2018)
[arXiv:1802.09983 [astro-ph.HE]].

\bibitem{Sui:2018bbh}
Y.~Sui and P.~S.~Bhupal Dev,
JCAP \textbf{07}, 020 (2018)
[arXiv:1804.04919 [hep-ph]].

\bibitem{Kachelriess:2018rty}
M.~Kachelriess, O.~E.~Kalashev and M.~Y.~Kuznetsov,
Phys. Rev. D \textbf{98}, no.8, 083016 (2018)
[arXiv:1805.04500 [astro-ph.HE]].

\bibitem{Chianese:2019kyl}
M.~Chianese, D.~F.~G.~Fiorillo, G.~Miele, S.~Morisi and O.~Pisanti,
JCAP \textbf{11}, 046 (2019)
[arXiv:1907.11222 [hep-ph]].

\bibitem{Ishiwata:2019aet}
K.~Ishiwata, O.~Macias, S.~Ando and M.~Arimoto,
JCAP \textbf{01}, 003 (2020)
[arXiv:1907.11671 [astro-ph.HE]].

\bibitem{Dekker:2019gpe}
A.~Dekker, M.~Chianese and S.~Ando,
JCAP \textbf{09}, 007 (2020)
[arXiv:1910.12917 [hep-ph]].

\bibitem{Neronov:2020wir}
A.~Neronov and D.~Semikoz,
Phys. Rev. D \textbf{102}, no.4, 043025 (2020)
[arXiv:2001.11881 [astro-ph.HE]].




\bibitem{Amenomori:2021gmk}
M.~Amenomori \textit{et al.} [Tibet ASgamma],
Phys. Rev. Lett. \textbf{126}, no.14, 141101 (2021)
[arXiv:2104.05181 [astro-ph.HE]].


\bibitem{He:2020ipy}
H.~He [LHAASO],
PoS \textbf{ICRC2019}, 693 (2021)

\bibitem{Fang:2021ylv}
K.~Fang and K.~Murase,
[arXiv:2104.09491 [astro-ph.HE]].

\bibitem{Liu:2021lxk}
R.~Y.~Liu and X.~Y.~Wang,
[arXiv:2104.05609 [astro-ph.HE]].

\bibitem{Qiao:2021iua}
B.~Q.~Qiao, W.~Liu, M.~J.~Zhao, X.~J.~Bi and Y.~Q.~Guo,
[arXiv:2104.03729 [astro-ph.HE]].

\bibitem{Dzhatdoev:2021xjh}
T.~A.~Dzhatdoev,
[arXiv:2104.02838 [astro-ph.HE]].

\bibitem{Lipari:2018gzn}
P.~Lipari and S.~Vernetto,
Phys. Rev. D \textbf{98}, no.4, 043003 (2018)
[arXiv:1804.10116 [astro-ph.HE]].

\bibitem{Bartoli:2015era}
B.~Bartoli \textit{et al.} [ARGO-YBJ],
Astrophys. J. \textbf{806}, 20
[arXiv:1507.06758 [astro-ph.IM]].

\bibitem{Borione:1997fy}
A.~Borione \textit{et al.}
Astrophys. J. \textbf{493}, 175-179 (1998)
[arXiv:astro-ph/9703063 [astro-ph]].

\bibitem{Bhattacharya:2019ucd}
A.~Bhattacharya, A.~Esmaili, S.~Palomares-Ruiz and I.~Sarcevic,
JCAP \textbf{05}, 051 (2019)
[arXiv:1903.12623 [hep-ph]].

\bibitem{Cohen:2016uyg}
T.~Cohen, K.~Murase, N.~L.~Rodd, B.~R.~Safdi and Y.~Soreq,
Phys. Rev. Lett. \textbf{119}, no.2, 021102 (2017)
[arXiv:1612.05638 [hep-ph]].

\bibitem{Sjostrand:2019zhc}
T.~Sj\"ostrand,
Comput. Phys. Commun. \textbf{246}, 106910 (2020)
[arXiv:1907.09874 [hep-ph]].

\bibitem{Bauer:2020jay}
C.~W.~Bauer, N.~L.~Rodd and B.~R.~Webber,
[arXiv:2007.15001 [hep-ph]].

\bibitem{Navarro:1996gj}
J.~F.~Navarro, C.~S.~Frenk and S.~D.~M.~White,
Astrophys. J. \textbf{490}, 493-508 (1997)
doi:10.1086/304888
[arXiv:astro-ph/9611107 [astro-ph]].

\bibitem{Sommers:2000us}
P.~Sommers,
Astropart. Phys. \textbf{14}, 271-286 (2001)
[arXiv:astro-ph/0004016 [astro-ph]].


\bibitem{Ahlers:2013xia}
M.~Ahlers and K.~Murase,
Phys. Rev. D \textbf{90}, no.2, 023010 (2014)
[arXiv:1309.4077 [astro-ph.HE]].

\end{thebibliography}
\end{document}